\documentclass[fleqn,usenatbib]{mnras}

\usepackage{newtxtext,newtxmath}
\usepackage[mathscr]{euscript}

\usepackage[T1]{fontenc}
\usepackage{ae,aecompl}
\usepackage[utf8]{inputenc}
\usepackage{ulem}

\DeclareRobustCommand{\VAN}[3]{#2}
\let\VANthebibliography\thebibliography
\def\thebibliography{\DeclareRobustCommand{\VAN}[3]{##3}\VANthebibliography}

\usepackage{graphicx}	% Including figure files
\usepackage{amsmath}	% Advanced maths commands

\usepackage{amssymb,bm}	% Extra maths symbols
\usepackage{xcolor}

\newcommand{\Reply}[1]{#1}
\title[Strangeon Stars]{Oscillation Modes and Gravitational Waves from Strangeon Stars}

\author[H.-B. Li et al.]{
Hong-Bo Li,$^{1,2}$
Yong Gao,$^{1,2}$\thanks{gaoyong.physics@pku.edu.cn}
Lijing Shao,$^{2,3}$\thanks{lshao@pku.edu.cn}
Ren-Xin Xu,$^{1,2}$\thanks{r.x.xu@pku.edu.cn}
and Rui Xu$^{2}$
\\
$^{1}$Department of Astronomy, School of Physics, Peking University, Beijing 100871, China\\
$^{2}$Kavli Institute for Astronomy and Astrophysics, Peking University, Beijing 100871, China\\
$^{3}$National Astronomical Observatories, Chinese Academy of Sciences, Beijing 100012, China
}

\date{Accepted XXX. Received YYY; in original form ZZZ}

\pubyear{2022}

\begin{document}
\label{firstpage}
\pagerange{\pageref{firstpage}--\pageref{lastpage}}
\maketitle

\begin{abstract}
The strong interaction at low energy scales determines the equation of state
(EOS) of supranuclear matters in neutron stars (NSs). It is conjectured that the
bulk dense matter may be composed of strangeons, which are quark clusters with
nearly equal numbers of $u$, $d$, and $s$ quarks. To characterize the
strong-repulsive interaction at short distance and the nonrelativistic nature of
strangeons, a phenomenological Lennard-Jones model with two  parameters is used to describe the EOS
of strangeon stars (SSs). For the first time, we investigate the oscillation
modes of non-rotating SSs and obtain their frequencies for various
parameterizations of the EOS.  We find that the properties of radial
oscillations of SSs are different from those of NSs, especially for stars with
relatively low central energy densities.  Moreover, we calculate the $f$-mode
frequency of nonradial oscillations of SSs within the relativistic Cowling
approximation. The frequencies of the $f$-mode of SSs are found to be in the
range from $6.7\,$kHz to $ 8.7\,\rm{kHz}$.  Finally, we study the universal
relations between the $f$-mode frequency and global properties of SSs, such as 
the compactness and the tidal deformability. The results we obtained are
relevant to pulsar timing and gravitational waves, and will help to probe NSs'
EOSs and infer nonperturbative behaviours in quantum chromodynamics.
\end{abstract}


%%%%%%%%%%%%%%%%%%%%%%%%%%%%%%%%%%%%%%%%%%%%%%%%%%
\allowdisplaybreaks  % allow long equations to display on different pages
%%%%%%%%%%%%%%%%% BODY OF PAPER %%%%%%%%%%%%%%%%%%
%---------------------------------------------------------------------
\section{Introduction}\label{sec:Introduction}
%---------------------------------------------------------------------

The equation of state (EOS) of nuclear dense matter plays a crucial role in many
astrophysical phenomena associated with neutron
stars~\citep[NSs;][]{Ozel:2010fw, Lattimer:2006xb, LIGOScientific:2018cki}.
Owing to the non-perturbative properties of the strong interaction at low
energy, the EOS of dense matters at several nuclear densities still remains
unknown.  \citet{Witten:1984rs} conjectured that the true ground state of the
dense matter is quark matter composed of almost free $u$, $d$, and $s$ quarks.
The pulsar-like compact objects should be quark stars (QSs) rather than
conventional NSs.  The MIT bag model with almost free
quarks~\citep{Alcock:1986hz} and the color-superconductivity state
model~\citep{Alford:2007xm} have been used in literature to study QSs.  In
2003,~\citet{Xu:2003xe} proposed that the constituting units of the supranuclear
matter could be strange quark clusters, since the non-perturbative strong
interaction may render quarks grouped in clusters.  Each quark cluster is
composed of several quarks (including $u$, $d$, and $s$ flavors) condensing in
position space rather than in momentum space. A name ``strangeon'' is coined to
these strange ``nucleons''~\citep{Xu:2016uod, Lai:2017ney}.  
In this sense, compressed baryonic
matter could be in a state of strangeons, and pulsar-like compact stars could thus be
strangeon stars (SSs).

Strangeon matter, similar to strange quark matter, is composed of nearly equal
numbers of $u$, $d$, and $s$ quarks. However, different from strange quark
matter, quarks in strangeon matter are localized inside strangeons due to the
strong coupling between quarks. There are differences and similarities among
NSs, QSs, and SSs. On the one hand, quarks are thought to be localized in
strangeons in SSs, like neutrons in NSs. On the other hand, a strangeon , with 
light-flavour symmetry restoration of quark, may contain more than three valence quarks. In addition, the
matter at the surface of SSs is strangeon matter, i.e., SSs are self-bound by
the strong force, like QSs~\citep{Xu:2003xe}. These properties are fundamental
to a few important astrophysical observables.  A sophisticated study on various
global parameters of rotating SSs, including mass, radius, moment of inertia,
tidal deformability, quadrupole moments, and shape parameters, was carried out
by \citet{Gao:2021uus}.

SSs can account for  many current observational facts in astrophysics.
The EOS of SSs could be very stiff to explain the observed massive
pulsars~\citep{Demorest:2010bx, Antoniadis:2013pzd}. The magnetospheric activity
of SSs was discussed in~\citet{Xu:1999bw}. \citet{FAST:2019zow}
explained the sub-pulse drifting of radio pulsars using the properties at the
surface of SSs. Also, pulsar glitches could be the result of star-quakes~\citep{ Peng:2007eq, Zhou:2004ue, Zhou:2014tba}, and a detailed modeling of
the glitch behaviours confronted with observations was discussed
in~\citet{Lai:2017xys}. The model of SSs can be extended to explain the glitch
activity of normal radio pulsars~\citep{Wang:2020xsm}. Recent
studies~\citep{Lai:2018ugk, Lai:2017mjv, Lai:2020mlu} have investigated the
tidal deformability as well as the ejecta and light curve of merging binary SSs,
showing consistency with the observations of the gravitational wave (GW) event
GW170817~\citep{LIGOScientific:2017vwq} and its multiwavelength electromagnetic
counterparts~\citep{Kasliwal:2017ngb, Kasen:2017sxr}. 

Owing to the difficulties in determining the EOS of pulsar-like compact stars
from first principles, observations from different channels become important
avenues in studying the EOS at high density, which can in turn be used to
constrain microscopic laws~\citep{Ozel:2010fw}.  In this respect, GW
asteroseismology that deals with oscillation modes offers a promising channel in
the new era of GWs~\citep{Andersson:1997rn, Benhar:2004xg, Doneva:2013zqa,
Andersson:2019}. It is the focus of this study.

Radial oscillations of stellar models were studied in the pioneering works
of~\citet{Chandrasekhar:1964zza, Chandrasekhar:1964zz}.  Notably, the properties
of radial oscillations can give information about the stability and the EOS of
compact stars.  The first exhaustive compilation of radial oscillations for
different zero-temperature EOSs was presented by~\citet{Glass:1983}.
In~\citet{Vaeth:1992}, the properties of radial modes of quark stars (QSs) were
investigated.  Furthermore, the study of radial oscillations of zero-temperature
NSs can be extended to proto-NSs~\citep{Gondek:1997fd}. Because the EOS of
proto-NSs is significantly softer than that of zero-temperature NSs, their
spectra of the radial oscillation modes are very different.  It is worth noting
that~\citet{Kokkotas:2000up} presented a useful survey on the radial oscillation
modes of NSs for various EOSs.  Based on the equations presented
by~\citet{Misner:1973}, they showed that the derivatives in the linear
differential equation can be written in the form of a self-adjoint differential
operator.  In this work, we calculate the frequencies of the first three radial
modes of SSs using the method of the self-adjoint differential operator
\citep{Kokkotas:2000up}. By that we can investigate the properties of radial
oscillations of SSs in detail and study the stability of SSs rigorously.

Nonradial oscillations of relativistic stars were studied in the pioneering work
of~\citet{Thorne:1967}.  The oscillation modes are damped out due to the
emission of GWs, so these oscillation modes are called quasinormal modes (QNMs).
For typical non-rotating relativistic fluid stars, QNMs are classified in polar
and axial categories.  The polar modes include the fundamental ($f$) modes,
pressure ($p$) modes, and gravity ($g$) modes.  The axial modes only have the
spacetime ($w$) modes, which are directly associated with the spacetime metric
and have no analogy in the Newtonian theory of stellar
pulsations~\citep{Kokkotas:1992xak}.  A detailed discussion about the
relativistic perturbation equations was given in many works \citep[see
e.g.,][]{Lindblom:1983ps, Detweiler:1985zz, Chandrasekhar:1991fi, Allen:1997xj,
Kokkotas:1999bd}. Using the Cowling approximation~\citep{Cowling:1941},
~\citet{Sotani:2010mx} calculated nonradial
oscillations of NSs with hadron-quark mixed phase transition.
Besides,~\citet{Doneva:2012rd} investigated nonradial oscillations of
anisotropic NSs with polytropic EOSs. In \citet{Das:2021dru}, the impact 
of the dark matter on the f-mode was also studied.

The $f$-mode of NSs, QSs, and SSs is important for several reasons: (i) it
depends on the EOS of compact stars; (ii) it is expected to be excited in many
astrophysical scenarios and leads to efficient GW emission; (iii) its frequency
is lower than other QNMs such as that of the $p$-modes and the $w$-modes, hence
the $f$-mode oscillation is most likely to be detectable with a third-generation
detector like the Einstein Telescope and the Cosmic
Explorer~\citep{Punturo:2010zz, Sathyaprakash:2019yqt, Kalogera:2021bya}, or
even in an optimal case by the current generation LIGO/Virgo/KAGRA
detectors~\citep{LIGOScientific:2018ehx, LIGOScientific:2022myk, KAGRA:2022qtq}.
In this work, we calculate the $f$-mode frequency of SSs in the Cowling
approximation and compare the results with those of NSs and QSs.

GW observation will be a powerful tool to study the EOS of compact stars in
particular in the case that we have good empirical formulas for the QNMs as
functions of stellar parameters.  Indeed, universal empirical formulas relating
the dynamical responses of a compact star under external perturbations---such as
the $f$-mode frequency, the tidal and rotational deformations---to its global
physical parameters---such as the mass, the radius, and the moment of
inertia---have been discovered for NSs and QSs. For example, the I-Love-Q
relations, discovered by~\citet{Yagi:2013awa, Yagi:2013bca}, relate the moment
of inertia $I$, the tidal deformability $\lambda$, and the spin induced
quadrupole moment $Q$. Another example is the relation for the $f$-mode
frequency, the moment of inertia, and the tidal deformability~\citep[see
e.g.,][]{Chan:2014kua}.  \citet{Sotani:2021kiw} have investigated various
universal relations between several oscillation modes and the tidal
deformability.  Inspired by works on the universal relations for single
NSs~\citep{Andersson:1997rn, Benhar:2004xg, Doneva:2013zqa, Yagi:2013awa,
Yagi:2013bca}, recent studies have investigated various universal relations
between the binary tidal deformability and the $f$-mode frequency of the
post-merger remnant of a binary NS system using numerical relativity
simulations~\citep{Bernuzzi:2015rla, Rezzolla:2016nxn, Kiuchi:2019kzt}.  
\Reply{It is worth noting that,~\citet{Kruger:2019zuz} and
\citet{Manoharan:2021wds} calculated the $f$-mode frequency for fast rotating
NSs without using the Cowling approximation, and discovered a relation between
the pre-merger tidal deformability and the dominant oscillation frequency (i.e.,
$f$-mode) of the post-merger remnant of a binary NS system.  Meanwhile, using
the universal relation of $f$-mode~\citep{Kruger:2019zuz, Manoharan:2021wds}, 
\citet{Volkel:2021gke} and \citet{Volkel:2022utc} studied the Bayesian inverse
problem of rotating NSs.}  In this work, we will study the universal relation
between the $f$-mode frequency and the tidal deformability of SSs, which will be
a useful input for comparisons among NSs, QSs, and SSs.

The paper is organized as follows. In Sec.~\ref{sec: EOS}, we introduce the EOS
of SSs and obtain the structure of non-rotating SSs. Based on the background
solutions, in Sec.~\ref{sec: radial oscillation}, we integrate the equations of
relativistic radial oscillations to determine the $f$-mode frequency for
different EOSs of SSs.  In Sec.~\ref{sec: nonradial oscillation}, we calculate
the frequency of nonradial $f$-mode and the tidal deformability of SSs. New fits of
universal relation between them are discussed.  Finally, we summarize our work in
Sec.~\ref{sec: conclusion}. 

Throughout the paper, we adopt geometric units with $c=G=1$, where $c$ and $G$
denote the speed of light and the gravitational constant respectively. The
metric signature is $(-\,,+\,,+\,,+)$.

%---------------------------------------------------------------------
\section{EQUATION OF STATE and structure of spherical static stars}\label{sec: EOS}
%---------------------------------------------------------------------

We assume that the interaction potential between two strangeons is described by
the Lennard-Jones potential~\citep{Jones:1924, Lai:2009cn, Gao:2021uus},
%--
\begin{equation}\label{LJ}
  u(r)=4 \epsilon\left[ \left(\frac{\sigma}{r}\right)^{12} -
  \left(\frac{\sigma}{r}\right)^{6} \right] \,,
\end{equation}
%--
where $\epsilon$ is the depth of the potential, $r$ is the distance between two
strangeons, and $\sigma$ is the distance when $u(r)=0$.  We note that this
potential has the property of short-distance repulsion and long-distance
attraction.

According to the results of early studies~\citep{Xu:2003xe, Lai:2009cn,
Gao:2021uus}, the potential energy density is given by
%--
\begin{equation}
    \rho_{\rm p}=2 \epsilon\left(A_{12} \sigma^{12} n^{5}-A_{6} \sigma^{6}
    n^{3}\right)\,,
\end{equation}
%--
where $A_{12}=6.2$, $A_{6}=8.4$, and $n$ is the number density of strangeons.
The total energy density of zero-temperature dense matter composed of strangeons
reads
%--
\begin{equation}\label{eq:energy_density}
    \rho=2 \epsilon\left(A_{12} \sigma^{12} n^{5}-A_{6} \sigma^{6} n^{3}\right)
    + nN_{\rm q}m_{\rm q}\,,
\end{equation}
%--
where $N_{\rm q}m_{\rm q}$ is the mass of a strangeon with $N_{\rm q}$ being the
number of quarks in a strangeon and $m_{\rm q}$ being the quark mass. In the
above equation the contributions from degenerate electrons and vibrations of the
lattice are neglected.  From the first law of thermodynamics, one derives the
pressure
%-----------------------------------
\begin{equation}\label{eq:pressure}
    P=n^{2} \frac{\mathrm{d}(\rho / n)}{\mathrm{d} n}=4 \epsilon\left(2 A_{12}
    \sigma^{12} n^{5}-A_{6} \sigma^{6} n^{3}\right)\,.
\end{equation}
%-----------------------------------
At the surface of  SSs, the pressure becomes zero and we obtain the surface
number density of strangeons as $\big[A_{6}/(2A_{12}\sigma^{6})\big]^{1/2}$. For
convenience, we transform it to the number density of baryons
%-----------------------------------
\begin{equation} \label{eq:surface}
    n_{\rm s}=\left(\frac{A_{6}}{2A_{12}}\right)^{1/2}\frac{N_{\rm
    q}}{3\sigma^{3}}\,.
\end{equation}
%-----------------------------------

%%%%%%%%%%%%%%%%%%%%%%%%%%%%%%%%%%%%%%%%%%%%%%%%%%
\begin{figure}
    \centering
    \includegraphics[width=8cm]{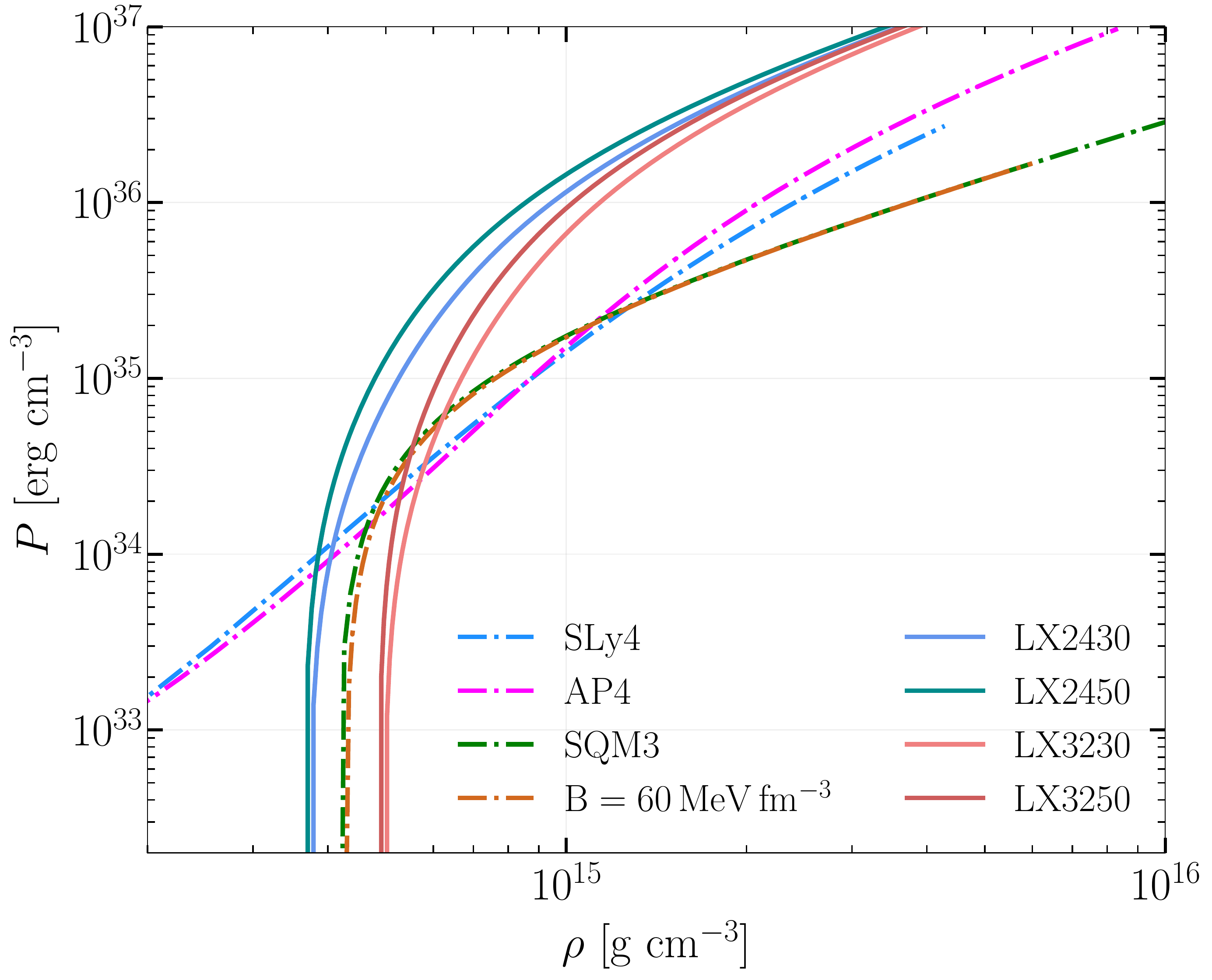}
    \caption{Relations between mass-energy density $\rho$ and 
    pressure $P$ for NSs, QSs, and SSs.}
    \label{fig:EOS_SSs}
\end{figure}
%%%%%%%%%%%%%%%%%%%%%%%%%%%%%%%%%%%%%%%%%%%%%%%%%%

For a given number of quarks $N_{\rm q}$ in a strangeon, the EOS of SSs is
completely determined by the depth of the potential $\epsilon$ and the number
density of baryons $n_{\rm s}$ at the surface of the star.  An 18-quark cluster,
called quark-alpha~\citep{Michel:1991pk}, can be completely symmetric in spin,
flavor, and color spaces. Therefore, we set $N_{\rm q}=18$ in our calculation as
a reasonable example.

\Reply{Besides the EOS of SSs, we also consider six EOSs of NSs and QSs for
comparison, including four popular nuclear matter EOSs for NSs,
AP4~\citep{Akmal:1997ft}, SLy4~\citep{Douchin:2001sv}, MS0, and
MS2~\citep{Mueller:1996pm}, as well as two QS models, the MIT bag model with a
bag constant $B=60\, \rm MeV \, fm^{-3}$~\citep{Alcock:1986hz} and
SQM3~\citep{Lattimer:2000nx}.\footnote{\Reply{Notice that, the nucleonic EOSs,
MS0 and MS2, have similar properties as SSs of higher maximum masses.}}} The corresponding
density-pressure relations for these EOSs are depicted in
Fig.~\ref{fig:EOS_SSs}.  We denote the EOSs of SSs using their values of $n_{\rm
s}$ and $\epsilon$. For example, ``LX2430'' means a surface baryon number
density $n_{\rm s}=0.24\,\rm fm^{-3}$ and a potential depth $\epsilon=30\,\rm
{MeV}$.

We consider the unperturbed relativistic star to be described by a perfect
fluid. The energy-momentum tensor is
$T_{\mu\nu} = (\rho + P)u_\mu u_\nu + P\,g_{\mu\nu}$.
The static and spherically symmetric metric, which describes an equilibrium
relativistic star, is given by the line element,
%-----------------------------------
\begin{equation}
  {\rm d} s^2 = -e^{2\Phi} {\rm d}t^2 + e^{2\Lambda} {\rm d} r^2 + r^2({\rm d}
  \theta^2 + \sin^2\theta {\rm d}\phi^2) \,,
\end{equation}
where $\Phi$ and $\Lambda$ are metric functions of $r$.
%-----------------------------------
A mass function $m(r)$ is defined as $m(r)=r(1-e^{-2\Lambda})/2$, which satisfies
%-----------------------------------
\begin{equation}\label{eq:dmdr}
 \frac{{\rm d} m}{{\rm d} r} = 4\pi r^2\rho \,,
\end{equation}
%-----------------------------------
where $\rho$ is the energy density.  The Tolman-Oppenheimer-Volkoff (TOV)
equations that determine the pressure $P(r)$ and the metric function $\Phi(r)$
are expressed as
%-----------------------------------
\begin{align}
\frac{{\rm d} P}{{\rm d} r} &= -(\rho+P)\frac{{\rm d} \Phi}{{\rm d} r}
\label{eq:dpdr} \,, \\
\frac{{\rm d} \Phi}{{\rm d} r} &= \frac{m+4\pi r^3 P}{r(r-2m)}  \label{eq:tov}
\,.
\end{align}
%-----------------------------------
Integrating Eqs.~(\ref{eq:dmdr}), ~(\ref{eq:dpdr})  and (\ref{eq:tov}) combined
with the EOS, one obtains the stellar structure of spherical stars and the
spacetime geometry.  \Reply{In Fig.~2, we show the mass-radius relations for
NSs, QSs, and SSs using the aforementioned EOSs. The EOSs of SSs are very stiff
because the strangeons are nonrelativistic and there is a very strong repulsion
at a short inter-cluster distance~\citep{Gao:2021uus}, which leads to the
maximal masses over $3\,M_{\odot}$. In contrast, the quarks are relativistic and
nearly free for QSs, so the EOSs are soft and the maximal masses only reach $2
\, M_{\odot}$ marginally.  The observations of the massive pulsars,
PSRs~J0348+0432~\citep{Antoniadis:2013pzd} and
J0740$+$6620~\citep{Fonseca:2021wxt}, at $\sim2 \,M_{\odot}$ via pulsar timing
support the stiff properties of the EOS.  More massive ones (e.g., $\geq 2.5 \,
M_{\odot}$) are expected in our model for future discovery.  The GWs from the
binary NS inspiral, GW170817, gave constraints on the tidal deformability for
the first time~\citep{LIGOScientific:2017vwq, LIGOScientific:2018cki,
LIGOScientific:2018hze}, which rules out  several stiff EOSs (e.g., EOSs MS0 and
MS2) and models of SSs with very low surface baryonic densities (say, LX2430 and
LX2450) at a $90\%$ credible level (see Fig.~18 in ~\citet{Gao:2021uus}).}

%%%%%%%%%%%%%%%%%%%%%%%%%
\begin{figure}
    \centering
   \includegraphics[width=8cm]{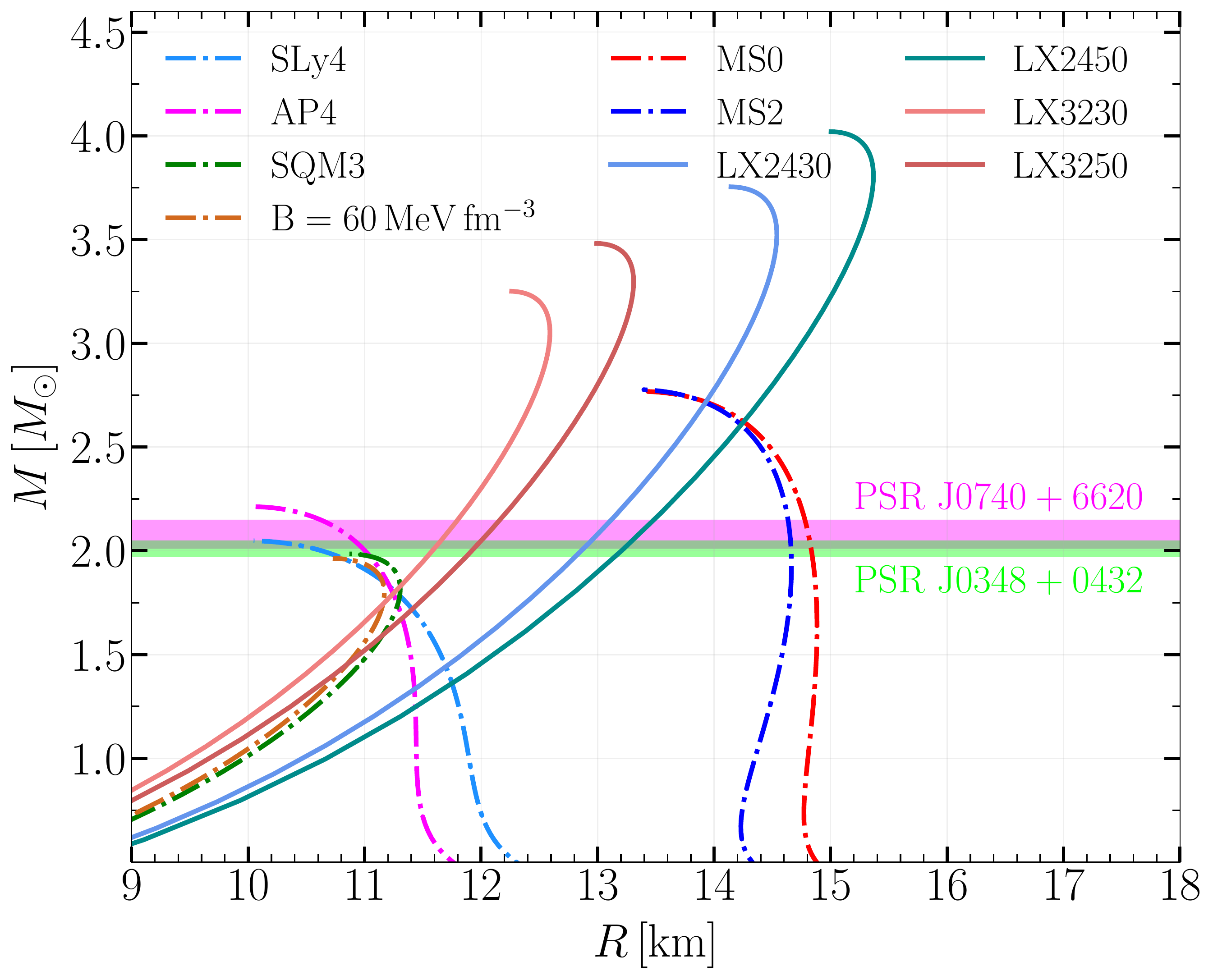}
    \caption{Mass-radius relations of SSs with different combinations of the
    surface baryonic density $n_{\rm s}$ and the potential depth $\epsilon$. For
    comparison, we also show the mass-radius relations for selected NSs and QSs.
    \Reply{The $1$-$\sigma$ regions of the mass measurements in
    PSRs~J0348+0432~\citep{Antoniadis:2013pzd} and
    J0740+6620~\citep{Fonseca:2021wxt} are illustrated.}}
    \label{fig:MR_SSs}
\end{figure}
%%%%%%%%%%%%%%%%%%%%%%%%%

%-----------------------------------------------------------
\section{RADIAL OSCILLATIONS}\label{sec: radial oscillation}
%----------------------------------------------------------- 

In this section, we study radial oscillations of SSs. We denote the radial
displacement of a fluid element as $\delta r(r,t)$ and its harmonic oscillation
mode with circular frequency $\omega$ as $\delta r(r,t) = X(r)e^{i\omega t}$.
To obtain the discrete set of oscillation frequencies of SSs, we adopt the
perturbation equations in~\citet{Kokkotas:2000up}. In practice we define a new
variable $ \zeta=r^2 e^{-\Phi}X$.  The master equation for radial oscillations
is expressed as 
%--
\begin{equation}\label{eq:radial-1}
       \frac{ {\rm d} }{{\rm d} r}\Big(\mathcal{P}\frac{{\rm d} \zeta}{{\rm d}
       r}\Big) + \big(\mathcal{Q} +\omega^2 \mathcal{W}\big)\zeta=0    \,,
\end{equation}
%--
where
%-----------------------------------
\begin{align}\label{eq:radial-2}
r^2 \mathcal{P} &= \Gamma P\,e^{(\Lambda + 3\Phi)}  \,,   \nonumber \\
r^2 \mathcal{Q} &= e^{(\Lambda + 3\Phi)}(\rho +P)\left[ (\Phi')^2+
4\frac{\Phi'}{r}- 8\pi e^{2\Lambda}P \right] \,,   \nonumber \\
r^2 \mathcal{W} &= (\rho + P)e^{(3\Lambda + \Phi)}  \,.
\end{align}
%-----------------------------------
By setting $\eta = \mathcal{P}\zeta'$,
%-----------------------------------
one obtains the following coupled differential equations,
%-----------------------------------
\begin{align}
 \frac{{\rm d} \zeta}{{\rm d} r} & =  \frac{ \eta}{\mathcal{P}}
 \label{eq:radial-4}  \,,  \\
 \frac{{\rm d} \eta}{{\rm d} r} & =  -\big(\omega^2\mathcal{W}+\mathcal{Q}\big)\zeta
 \label{eq:radial-5}  \,.
\end{align}
%-----------------------------------
At the center of the star, the boundary condition is 
$3\zeta_{0}={\eta_{0}}/{\mathcal{P}_{0}}$,  where $\zeta_{0}$ and $\eta_{0}$ are the values of $\zeta$
and $\eta$ at $r=0$ respectively \citep{Kokkotas:2000up}.  By setting
$\eta_{0}=1$, we have $\zeta_{0}={1} / {3\mathcal{P}_{0}}$, 
\Reply{where $\mathcal{P}_{0}= \Gamma P(0)e^{(\Lambda(0) + 3\Phi(0))}$}.  At the star surface
$r=R$, the pressure perturbation must vanish,  namely $ \Delta P =0 $, which
provides another boundary condition, $ \Gamma P \zeta' =0$.
Equations~(\ref{eq:radial-4}) and (\ref{eq:radial-5}) with the above two
boundary conditions form a two-point boundary value problem of the
Sturm-Liouville type with eigenvalues $\omega_{0}^2<\omega_{1}^2<\omega_{2}^2<
\cdots$~\citep{Shapiro:1983}, where $\omega_{0}$ is the eigenfrequency of the
$f$-mode.  If $\omega_{0}^2>0$, all the eigenfrequencies of the oscillation
modes are real, which indicates that the equilibrium stellar model is
dynamically stable~\citep{Chandrasekhar:1964zza, Chandrasekhar:1964zz,
Misner:1973}.  The period of the $f$-mode is given by
$\tau_{0}=1/\nu_{0}=2\pi/\omega_{0}$, where $\nu_{0}$ is the ordinary or
temporal frequency.  Inversely, $\omega_{0}^2<0$ corresponds to an exponentially
growing unstable radial oscillation. 

\begin{figure}
    \centering
    \includegraphics[width=8cm]{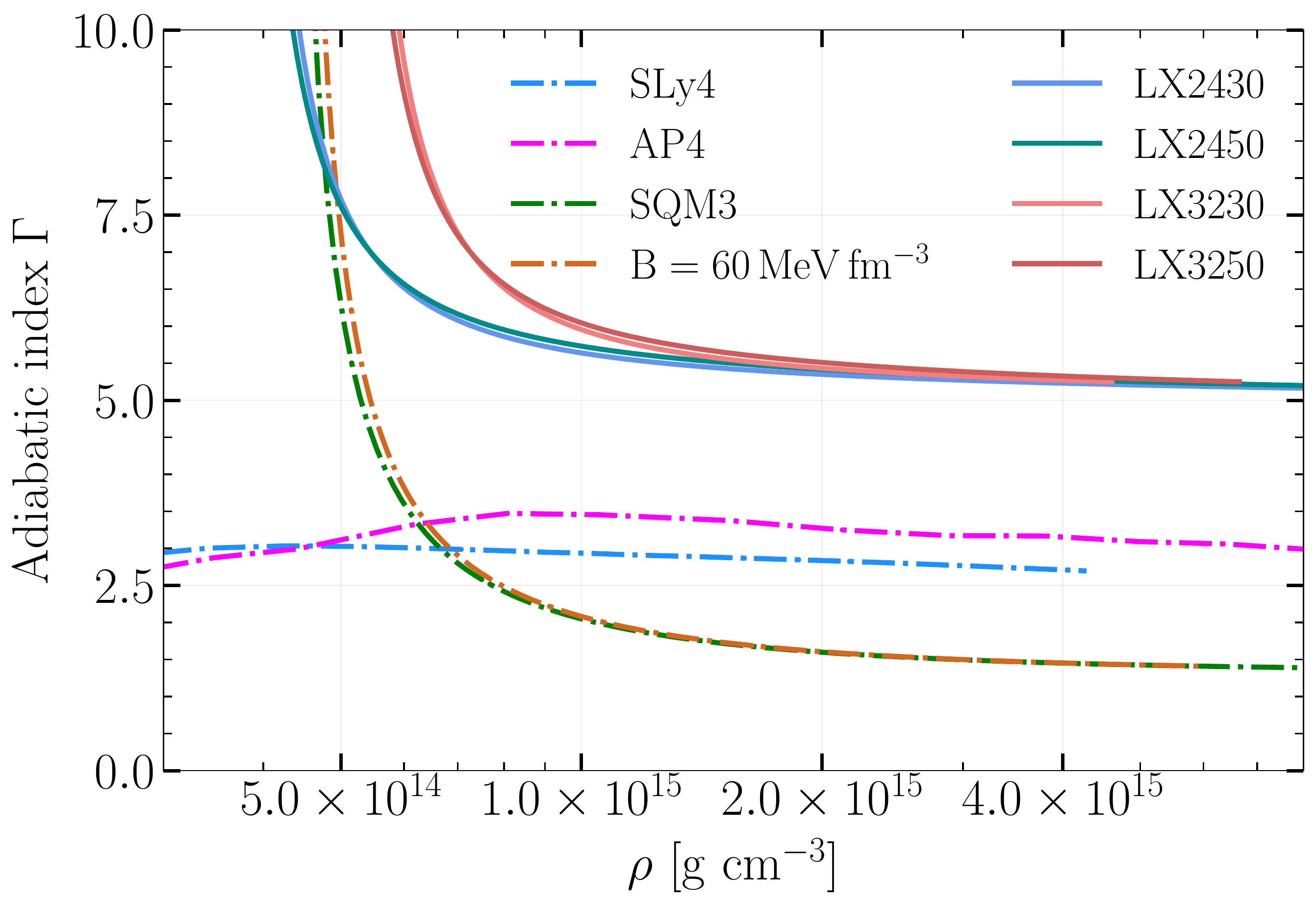}
    \caption{Relation between the adiabatic index $\Gamma$ and the
    mass-energy density $\rho$ for NSs, QSs and SSs.}
    \label{fig:Gamma_SSs}
\end{figure}

For adiabatic oscillations, the adiabatic index governing the perturbations is
defined by~\citep{Kokkotas:2000up}
%-----------------------------------
\begin{equation}\label{eq:adiabatic index}
  \Gamma = \frac{\rho+P}{P} \frac{{\rm
  d} P}{{\rm d} \rho} \, ,
\end{equation}
%-----------------------------------
which is equal to the adiabatic index governing the equilibrium pressure-energy
density relation.  The relation between the adiabatic index $\Gamma$ and the
mass-energy density $\rho$ is shown in Fig.~\ref{fig:Gamma_SSs}.  We note that
the adiabatic indices for QSs and SSs are qualitatively different from that of
NSs at low density. Moreover, SSs generally have a larger adiabatic index than
NSs and QSs, indicating that the EOSs of SSs are stiffer~\citep{Gao:2021uus}.

%%%%%%%%%%%%%%%%%%%%%%%%%%%%%%%%%%%%%%%%%%%%%%%%%%
%%%%%%%%%%%%%%%%%%%%%%%%%%%%%%%%%%%%%%%%%%%%%%%%%%

%%%%%%%%%%%%%%%%%%%%%%%%%%%%%%%%%%%%%%%%%%%%%%%%%%
\begin{figure}
    \centering
    \includegraphics[width=8cm]{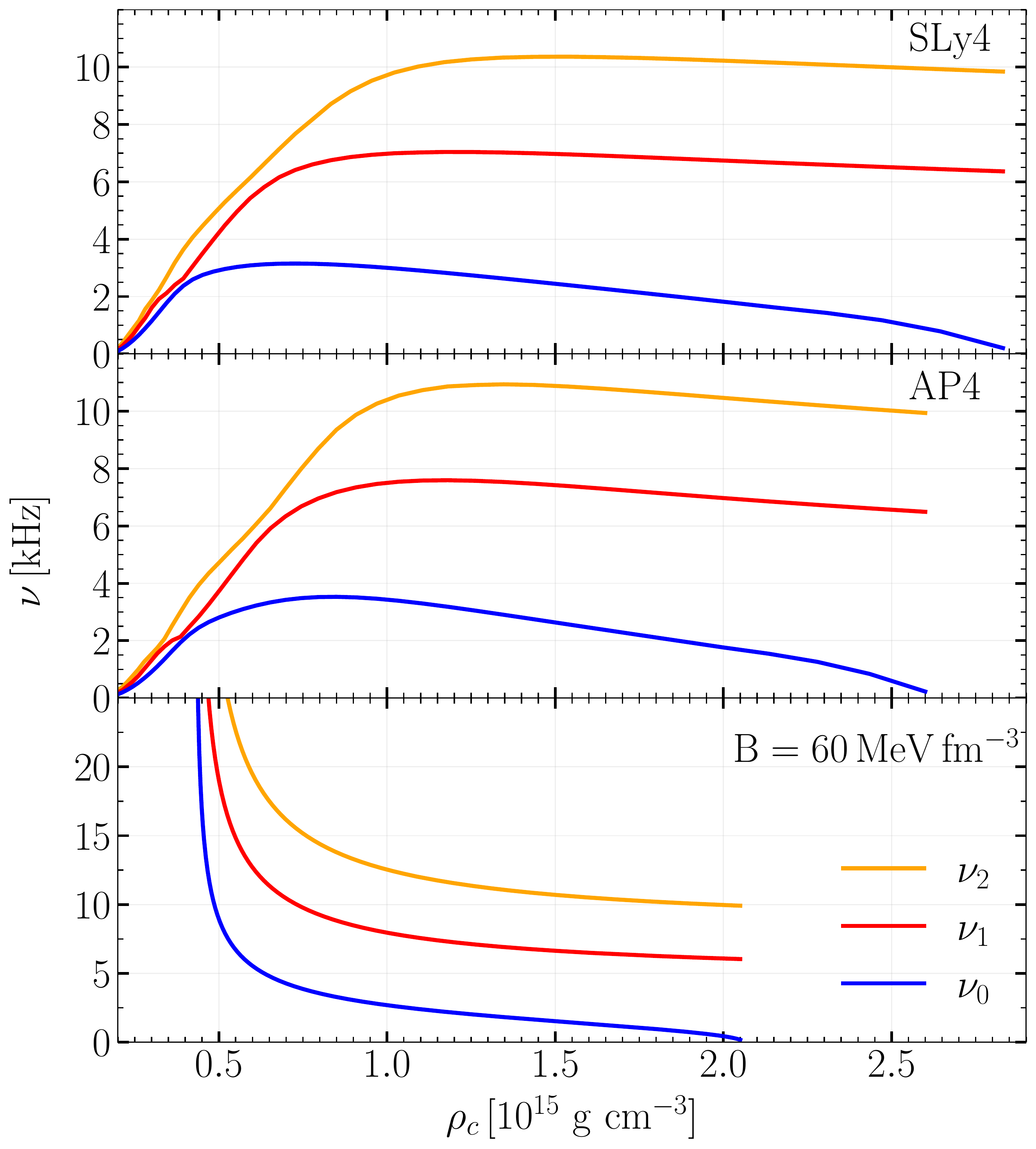}
    \caption{The frequencies of the fundamental mode, $\nu_{0}$, and the first
    two excited modes, $\nu_{1}$ and $\nu_{2}$, of radial oscillation, as
    functions of the central density $\rho_{c}$ for NSs and QSs.}
    \label{fig:radial_fd}
\end{figure}
%%%%%%%%%%%%%%%%%%%%%%%%%%%%%%%%%%%%%%%%%%%%%%%%%%

In Fig.~\ref{fig:radial_fd}, we present the $f$-mode frequency $\nu_{0}$ and the
frequencies of the first two excited modes, $\nu_{1}$ and $\nu_{2}$, for SLy4,
AP4, and the MIT bag model.  Our results for NSs reproduce the results
of~\citet{Kokkotas:2000up}.  We observe that $f$-mode becomes unstable (i.e.,
$\omega^2_0$ becoming negative) for central densities above
$2.83\times10^{15}\rm { g \, cm^{-3}}$, $2.70\times10 ^{15} \rm { g \,
cm^{-3}}$, and $2.05\times10 ^{15} \rm { g \, cm^{-3}}$ for three EOSs. The
instability point corresponds to maximal masses $2.04\, M_{\odot}$, $2.21\,
M_{\odot}$, and $1.96\, M_{\odot}$ for SLy4, AP4, and the MIT bag model
respectively.  It is worth noting that the $f$-mode frequency of the MIT bag
model behaves very different from that of NSs at low central density, rooting in
the self-bound and gravity-bound nature of QSs and NSs respectively.

To further explore the results for QSs, SSs, and NSs, we note that with a low
central density, the star can be approximated as a homogeneous nonrelativistic
star~\citep{Shapiro:1983}, so that the  angular frequency $ \omega_{0}$ of the
$f$-mode reads $\omega_{0}^{2}=4\pi \rho(4\Gamma-3) /3$.  Using the relations
between the density and the adiabatic index shown in Fig.~\ref{fig:Gamma_SSs},
we do expect the frequency $\omega_{0}$ to diverge as the density approaches a
minimal value for QSs and SSs.  For NSs, the adiabatic index does not change
significantly as the density decreases. Therefore for NSs, $\omega_{0}$ tends to
zero mildly when the central density of the star is sufficiently low.  Indeed,
these points are confirmed in Fig.~\ref{fig:radial_fd}. 

\begin{figure}
  \centering
  \includegraphics[width=8cm]{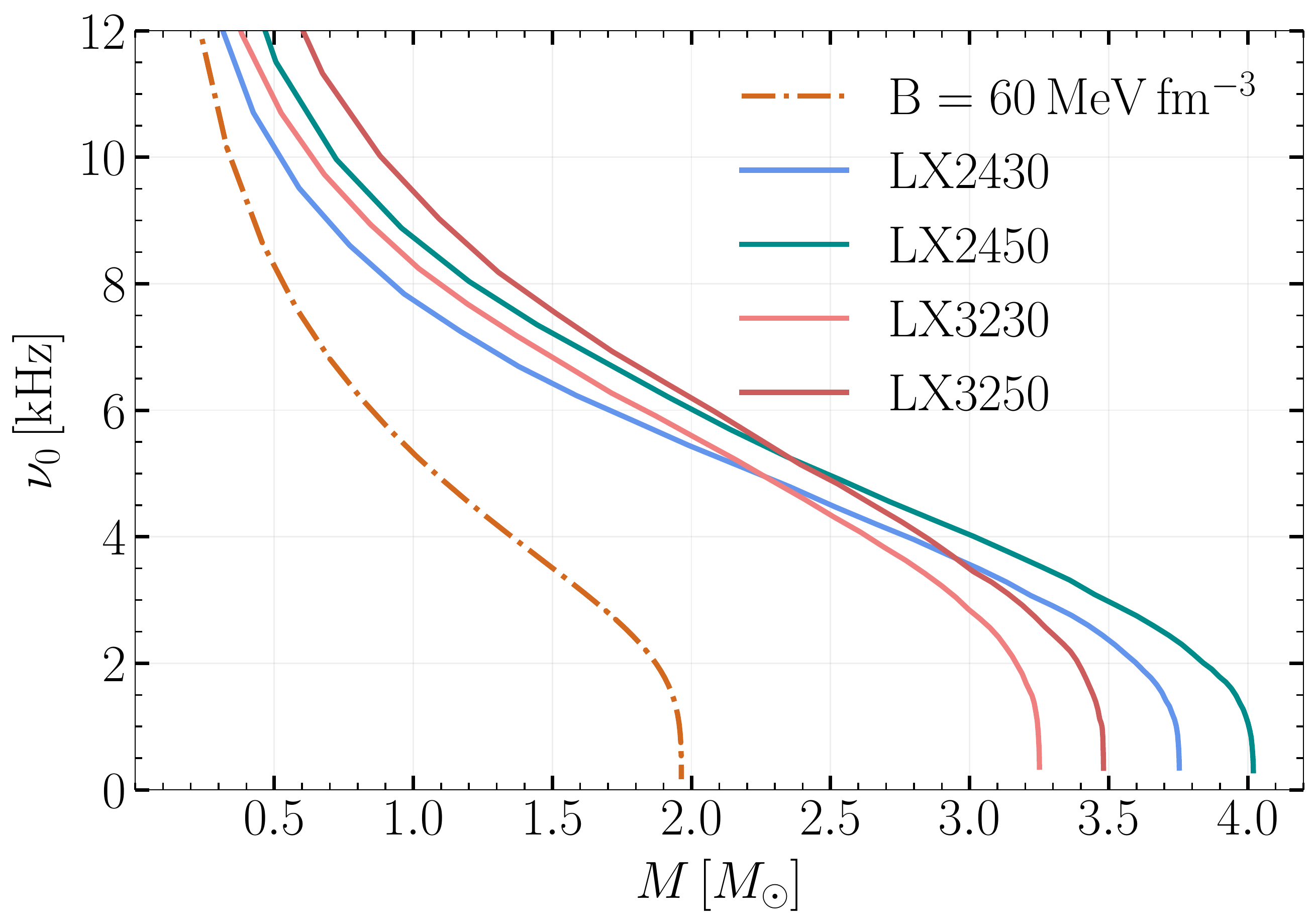}
  \caption{The frequencies of the fundamental mode $\nu_{0}$ as functions of the
  mass $M$ for QSs and SSs.}
  \label{fig:f_SSs}
\end{figure}

In Fig.~\ref{fig:f_SSs}, we show the ordinary frequency $\nu_{0}$ of the
$f$-mode versus the mass of the stars for SSs and one EOS of QSs.  The curves of
SSs have the same trend as that of QSs, with $\nu_{0}$ going to zero at their
maximal masses. However, $\nu_{0}$ for SSs is larger than that of QSs for a
given mass, which arises from the fact that SSs' EOSs are much stiffer than that
of QSs.

%%%%%%%%%%%%%%%%%%%%%%%%%%%%%%%%%%%%%%%%%%%%%%%%%%
%%%%%%%%%%%%%%%%%%%%%%%%%%%%%%%%%%%%%%%%%%%%%%%%%%

%---------------------------------------------------------------------------------
\section{nonradial OSCILLATIONS}\label{sec: nonradial oscillation}
%---------------------------------------------------------------------------------

In this section, we study nonradial oscillations of a non-rotating SS in the
Cowling approximation, in which the spacetime metric is kept to be the static
spherical background solution in the so-called Cowling
approximation~\citep{Cowling:1941}.  The fluid Lagrangian displacement vector is
given by
%-----------------------------------
\begin{equation}
  \label{eq:nonradialGR-1}
  \xi^i = \left(e^{-\Lambda}W, -V\partial_\theta,
  -V\sin^{-2}\theta\partial_\phi\right)r^{-2}Y_{\ell m} \,,
\end{equation}
%-----------------------------------
where $W$ and $V$ are functions of $t$ and $r$, while $Y_{\ell m}$ is the
spherical harmonic function.  Then the perturbation of the four-velocity,
$\delta u^\mu$, can be written as
%-----------------------------------
\begin{equation}\label{eq:nonradialGR-2}
  \delta u^\mu = \left(0, e^{-\Lambda}\partial_t W, -\partial_t V
  \partial_\theta, -\partial_t
  V\sin^{-2}\theta\partial_\phi\right)r^{-2}e^{-\Phi}Y_{\ell m}\,.
\end{equation}
%-----------------------------------

Assuming a harmonic dependence on time, the perturbative variables can be
written as $W(t,r)=W(r)e^{i\omega t}$ and $V(t,r) = V(r)e^{i\omega t}$.  We can
obtain the following system of equations for the fluid
perturbations~\citep[see][for a detailed variational derivation]{Sotani:2010mx,
Doneva:2012rd, Yazadjiev:2011sd},
%-----------------------------------
\begin{align}
  \frac{{\rm d} W}{{\rm d} r} &= \frac{{\rm d} \rho}{{\rm d} P}\left[\omega^2
  r^2e^{\Lambda-2\Phi}V +\frac{{\rm d} \Phi}{{\rm d} r} W\right] -
  \ell(\ell+1)e^{\Lambda}V  \,, \label{eq1} \\
  \frac{{\rm d} V}{{\rm d} r}  &= 2\frac{{\rm d} \Phi}{{\rm d} r}
  V-e^\Lambda\frac{W}{r^2} \,. \label{eq2}
\end{align}
%-----------------------------------

The boundary condition at the center of the star can be parameterized as,
$W=Ar^{l+1}$ and  $V=- {A} r^l / {l}$,
with $A$ being an arbitrary constant. It can be obtained by examining the
behavior of $W$ and $V$ in the vicinity of $r=0$.  At the surface of the star,
the perturbed pressure must vanish, which provides 
%-----------------------------------
\begin{equation}\label{eq:bc2}
    \omega^2 e^{\Lambda (R)-2\Phi (R)}V (R)+\frac{1}{R^2}\frac{{\rm d}\Phi}{{\rm
    d}r}\Big|_{r=R}W (R)=0 \,.
\end{equation}
%-----------------------------------

In full general relativity,  each QNM is characterized by a complex eigenfrequency
$\omega=\omega_{\rm r}+\rm {i}\,\omega_{\rm {i}}$~\citep{Thorne:1967}. The real
part $\omega_{\rm r}$ corresponds to the mode frequency, and the imaginary part
$\omega_{\rm {i}}$ gives the damping time $\tau\equiv1/\omega_{\rm {i}}$ due to
GW emission. However, in the Cowling approximation, we obtain normal modes of
oscillation and there is no emission of GWs.  For a non-rotating stellar model,
the Cowling approximation leads to a relative error $\sim 10\%$--$30\%$ for the
$f$-mode~\citep{Chirenti:2015dda, Sotani:2021jqa}. For higher modes, the
relative error is smaller~\citep{Yoshida:1997bf}.
 
%------------------------------------------------------------------------------------
\subsection{F-MODE FREQUENCY}\label{sec: different observablee}
%------------------------------------------------------------------------------------

Now we calculate the ordinary frequency $\nu_{0}$ of the $f$-mode for the $l=2$
nonradial oscillation, and study its relation with the mass $M$, the compactness
$C=M/R$, and the dimensionless tidal deformability $\Lambda$ for NSs, QSs, and
SSs. 

%%%%%%%%%%%%%%%%%%%%%%%%%%%%%%%%%%%%%%%%%%%%%%%%%%
\begin{figure}
    \centering
    \includegraphics[width=8cm]{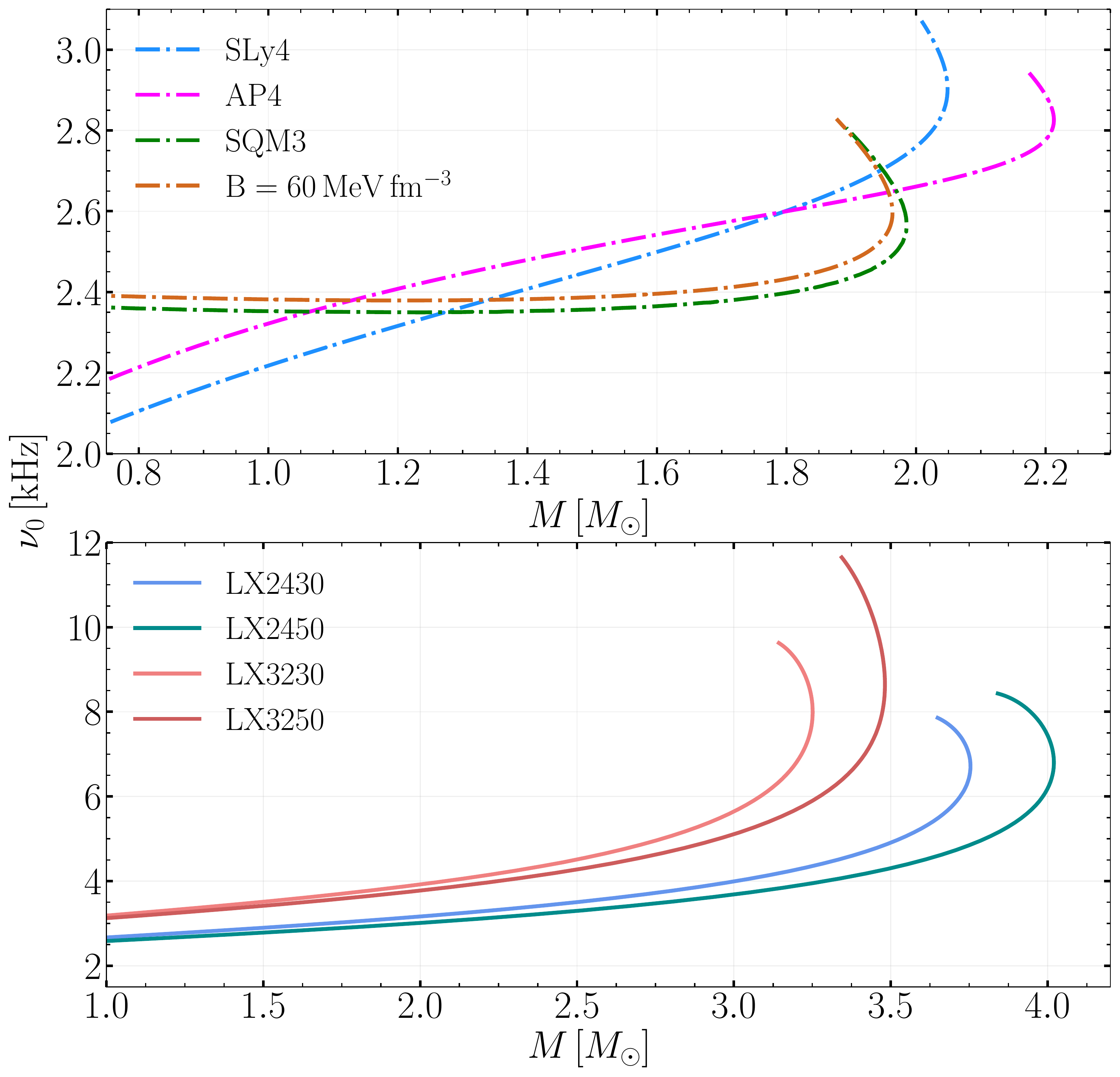}
    \caption{The frequency $\nu_{0}$ of $f$-mode as a function of mass $M$ for
     NSs and QSs (top), and  SSs (bottom).}
    \label{fig:fm}
\end{figure}
%%%%%%%%%%%%%%%%%%%%%%%%%%%%%%%%%%%%%%%%%%%%%%%%%%

The frequency $\nu_{0}$ versus mass $M$ for  NSs and QSs is shown in the top
panel of Fig.~\ref{fig:fm}.  By increasing the mass of the star, the frequency
$\nu_{0}$ increases significantly for NSs, while it does not change much for
QSs. This can be understood by noticing that QSs are self-bound by strong
interaction and the density in the interior of the star does not change too much
as the mass increases.  This is in contrast to NSs which are gravitationally
bound. From the figure, we can see that the values of $\nu_{0}$ at the maximal
masses of NSs and QSs are $2.907$\,kHz, $2.823$\,kHz, $2.562$\,kHz, and $2.597\,
\rm{kHz}$ for SLy4, AP4, SQM3, and the MIT bag model with $\rm B=60\,
MeV\,fm^{-3}$ respectively. 

Additionally, the frequency $\nu_{0}$ versus mass $M$ for SSs is shown in the
bottom panel of Fig.~\ref{fig:fm}. We find that the curves of SSs are similar to
those of QSs only that the frequency $\nu_{0}$ for SSs extends a much wider
range.  The values of $\nu_{0}$ at the maximal masses of SSs are $6.676$\,kHz,
$6.832$\,kHz, $7.977$\,kHz, and $8.684\, \rm{kHz}$ for the EOSs of SSs with
different values of $n_{\rm s}$ and $\epsilon$ that we use in the figure.
Compared with QSs and NSs, these values are much larger, and it could be an
indicator to distinguish EOSs via GW observations.

%%%%%%%%%%%%%%%%%%%%%%%%%%%%%%%%%%%%%%%%%%%%%%%%%%
\begin{figure}
    \centering
    \includegraphics[width=8cm]{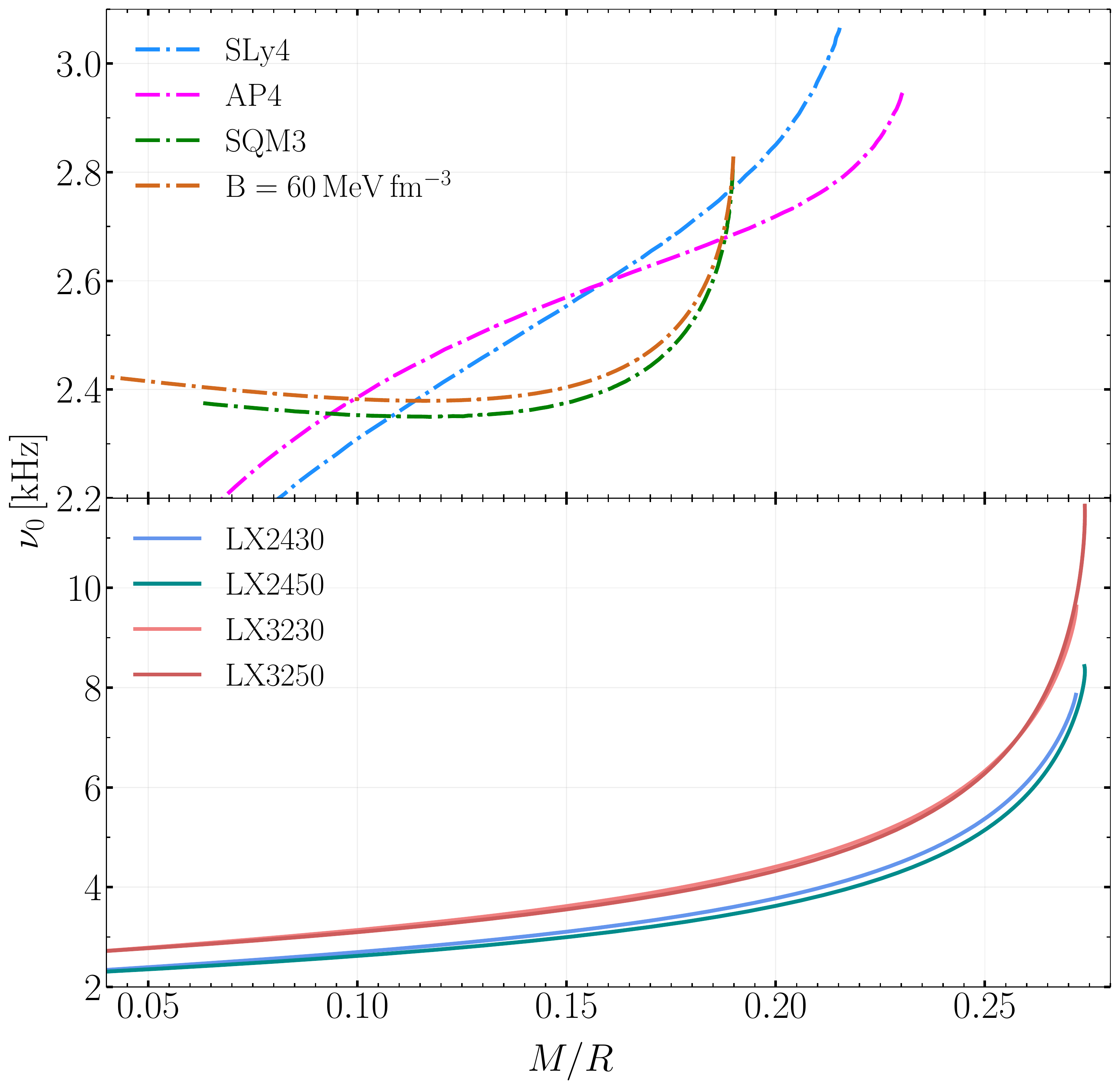}
    \caption{The frequency $\nu_{0}$ of the $f$-mode as a function of the
    compactness,  $C=M/R$, for NSs and QSs (top), and SSs (bottom).}
    \label{fig:fc}
\end{figure}
%%%%%%%%%%%%%%%%%%%%%%%%%%%%%%%%%%%%%%%%%%%%%%%%%%

We show in Fig.~\ref{fig:fc} the relation between the frequency $\nu_{0}$ and
the compactness of the stars.  It might be useful to note that the values of the
maximal compactness, $C_{\rm{max}}$, are $0.21$, $0.23$, $0.19$, and $0.19$ for
SLy4, AP4, SQM3, and the MIT bag model with $\rm B=60\, MeV \,fm^{-3}$,
respectively. In contrast, the value of $C_{\rm{max}}$ for SSs with different
values of $n_{\rm s}$ and $\epsilon$ is about the same, $C_{\rm{max}} \simeq
0.27$.  This maximal value of the compactness represents the limit of how stiff
EOSs of SSs can be due to the repulsive hardcore and the nonrelativistic nature
of strangeons.

%------------------------------------------------------------------------------------
\subsection{UNIVERSAL RELATIONS}\label{sec: universal relation}
%------------------------------------------------------------------------------------

To reveal the internal characters of NSs and assist relevant data analysis,
universal relations between the $f$-mode, $p$-mode, and $w$-mode frequencies and
the mass or the radius of NSs have been investigated~\citep{Andersson:1996pn,
Andersson:1997rn, Benhar:1998au, Benhar:2004xg, Tsui:2004qd}.  Motivated by
possible observations of the moment of inertia $I$ of NSs, \citet{Lau:2009bu}
used the moment of inertia to replace the compactness and discovered
EOS-independent relations in QNMs of NSs and QSs. Similar results were shown
in~\citet{Chirenti:2015dda}.  These relations can be used to infer the stellar
parameters---mass, radius, and possibly the EOS---from  QNM data with future GW
detectors.

%%%%%%%%%%%%%%%%%%%%%%%%%%%%%%%%%%%%%%%%%%%%%%%%%%
\begin{figure}
    \centering
    \includegraphics[width=8cm]{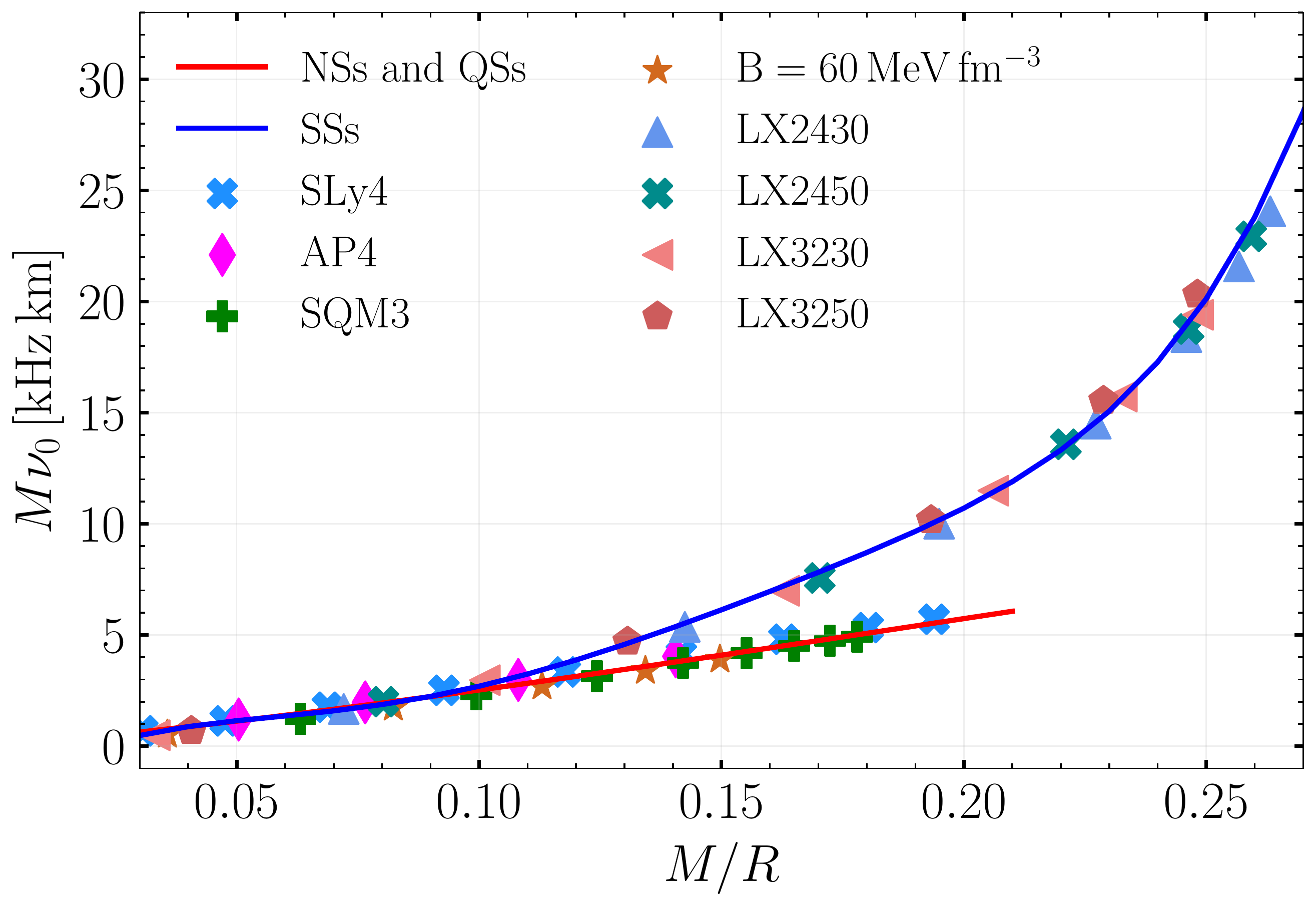}
    \caption{Scaled frequency of the $f$-mode as a function of the compactness
    $C$. The solid lines represent universal relations in Eqs.~(\ref{eq:
    fmc_NSs}) and ~(\ref{eq: fmc_SSs}).}
    \label{fig:relation_1}
\end{figure}
%%%%%%%%%%%%%%%%%%%%%%%%%%%%%%%%%%%%%%%%%%%%%%%%%%

Using the Cowling approximation,~\citet{Sotani:2010mx} calculated nonradial
oscillations of NSs with hadron-quark mixed phase transition, and discovered an
approximate formula. Inspired by the universal relation between the $f$-mode and
the compactness $C$~\citep{Sotani:2010mx}, we show the scaled frequency of the
$f$-mode versus the compactness $C$ for NSs, QSs, and SSs in
Fig.~\ref{fig:relation_1}.  In particular, as shown by the solid lines in the
figure, the universal relation for NSs can be represented by the following
empirical formula,
%--
\begin{equation}\label{eq: fmc_NSs}
  M \nu_{0}  = a_{\rm I} +b_{\rm I} \Big(\frac{M}{R}\Big)+c_{\rm I}
  \Big(\frac{M}{R}\Big)^{2} +d_{\rm I} \Big(\frac{M}{R}\Big)^{3} \,,
\end{equation}
%--
with $a_{\rm I}=-0.012$, $b_{\rm I}=19.48$, $c_{\rm I}=71.3$, and $d_{\rm
I}=-125$, while for SSs, we found a new universal relation,
%--
\begin{equation}\label{eq: fmc_SSs}
 M \nu_{0}= a_{\rm II} +b_{\rm II} \Big(\frac{M}{R} \Big)+c_{\rm II}
 \Big(\frac{M}{R} \Big)^{2} + d_{\rm II} \Big(\frac{M}{R} \Big)^{3} + e_{\rm II}
 \Big(\frac{M}{R} \Big)^{4} +  k_{\rm II}  \Big(\frac{M}{R} \Big)^{5}  \,,
\end{equation}
%--
with $a_{\rm II}=-2.795$, $b_{\rm II}=1.941\times10^{2}$, $c_{\rm
II}=-3.834\times10^{3}$, $d_{\rm II}=3.789\times 10^{4}$, $e_{\rm
II}=-1.601\times10^{5}$, and $k_{\rm II}=2.531\times10^{5}$.  We can observe
that the behavior of the $f$-mode frequencies for the SSs is very different from
the NSs and QSs, especially when the compactness is larger than $\sim 0.15$,
where the $f$-mode frequency from SSs is much larger than that of QSs and NSs.
It can be an important ``smoking gun'' signal for SSs.

For tidally deformed relativistic stars, the quadrupole tidal deformability
gives important information about the stellar structure.  To characterize the
deformation of the star, one usually defines the tidal deformability via $Q_{i
j} \equiv - \lambda \mathcal{E}_{i j}$, where $\mathcal{E}_{i j}$ is the
external tidal field and $Q_{i j}$ is the induced traceless quadrupole moment
tensor of stars~\citep{Hinderer:2007mb, Hinderer:2009ca}.  The parameter
$\lambda$ is related to the $l=2$ Love number $k_{2}$ via $k_{2}=3\lambda
R^{-5}/2$.  Besides,  the dimensionless tidal deformability $\Lambda$, defined
as $\Lambda =  {2}k_{2}C^{-5} / {3}$, is also commonly used.  We note that the
tidal deformability is proportional to the fifth power of the radius $R$.
Therefore, constraining or measuring tidal deformability can provide important
information on the EOS~\citep{LIGOScientific:2018cki, LIGOScientific:2018hze},
as well as test gravity theories~\citep{Hu:2021tyw, Xu:2021kfh}.  The influence
of tidal on the phase of GW in the inspiral stage is predominantly dependent on
the Love number $k_{2}$, and the effect enters at the fifth PN
order~\citep{Flanagan:2007ix}.

%%%%%%%%%%%%%%%%%%%%%%%%%%%%%%%%%%%%%%%%%%%%%%%%%%
\begin{figure}
    \centering
    \includegraphics[width=8cm]{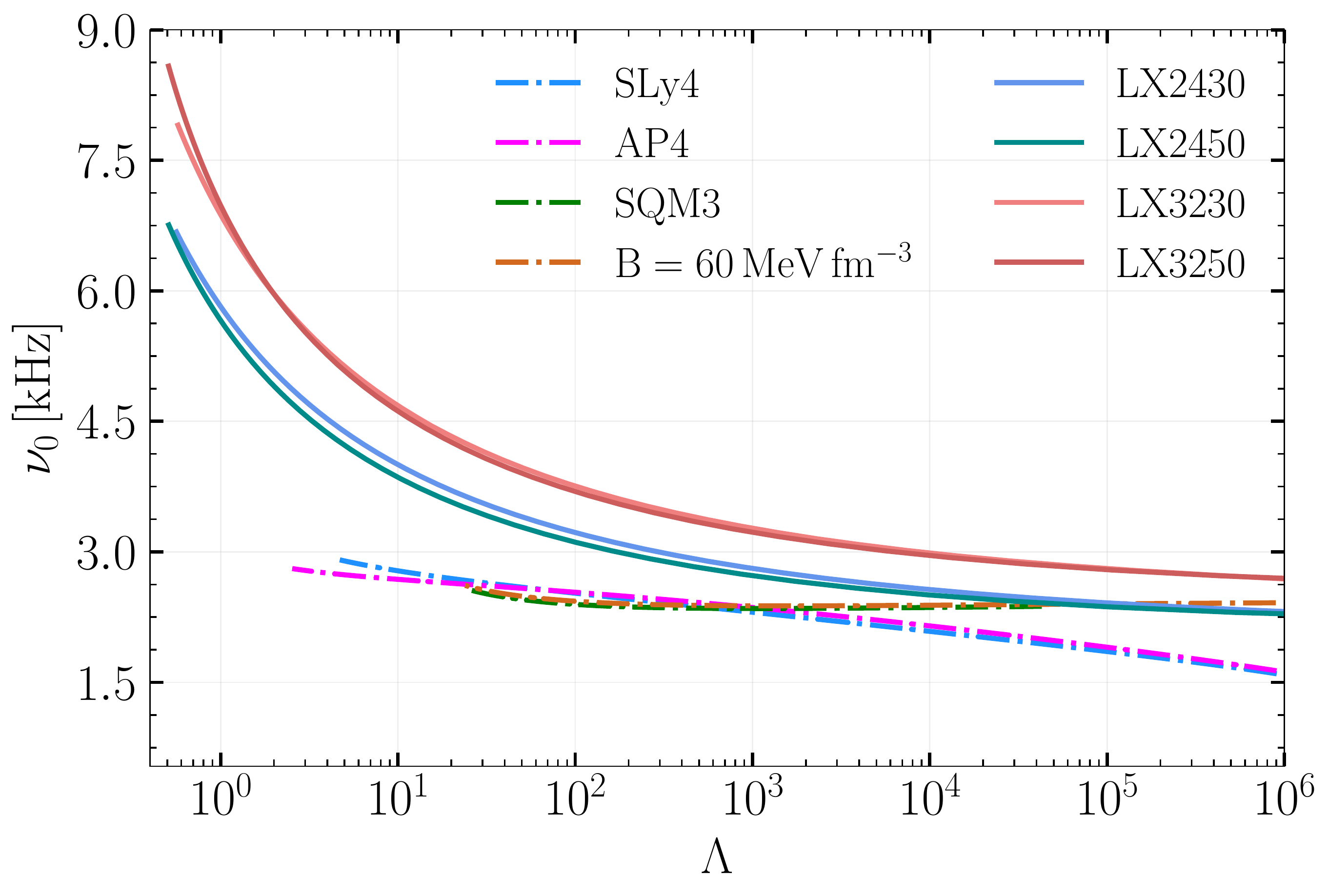}
    \caption{The frequency of the $f$-mode as a function of the dimensionless
    tidal deformability $\Lambda$ for NSs, QSs and SSs.}
    \label{fig:relation_2}
\end{figure}
%%%%%%%%%%%%%%%%%%%%%%%%%%%%%%%%%%%%%%%%%%%%%%%%%%

In Fig.~\ref{fig:relation_2}, we display the relation between the frequency of
the $f$-mode and the dimensionless tidal deformability, $ \Lambda$,  for NSs,
QSs, and SSs.  It is seen that the frequency decreases with $\Lambda$.  It is
understood that the more compact the star becomes the harder it can be deformed.
For SSs, as the potential depth $\epsilon$ increases and the surface baryonic
density $n_{\rm s}$ decreases, the EOS becomes stiffer, which leads to larger
tidal deformability and $f$-mode frequency.

For binary NSs of masses $M_{a}$ and $M_{b}$,  the dimensionless tidal coupling
constant is defined as~\citep{Bernuzzi:2014kca, Bernuzzi:2014owa,
Bernuzzi:2015rla},
%--
\begin{equation}\label{tidal}
  \kappa_{2}^{t} = 2\left[  q \left(\frac{X_{a}}{C_{a}}\right)^{5}k_{2}^{a}
  +\frac{1}{q}\left(\frac{X_{b}}{C_{b}}\right)^{5} k_{2}^{b} \right] \,,
\end{equation}
%--
where $q=M_{b}/M_{a} \leq1$, $X_{a} = M_{a}/(M_{a}+M_{b})$, and $C_i$ and
$k^i_2$ ($i=a,b$) are the compactness and the quadrupole Love number of each
star. If we consider a
binary system with non-rotating equal-mass configuration, the dimensionless
tidal coupling constant is given by
$ \kappa_{2}^{t} = {k_{2}} /{8C^{5}}= {3}\Lambda / 16$.

%%%%%%%%%%%%%%%%%%%%%%%%%%%%%%%%%%%%%%%%%%%%%%%%%%
\begin{figure}
    \centering
    \includegraphics[width=8.4cm]{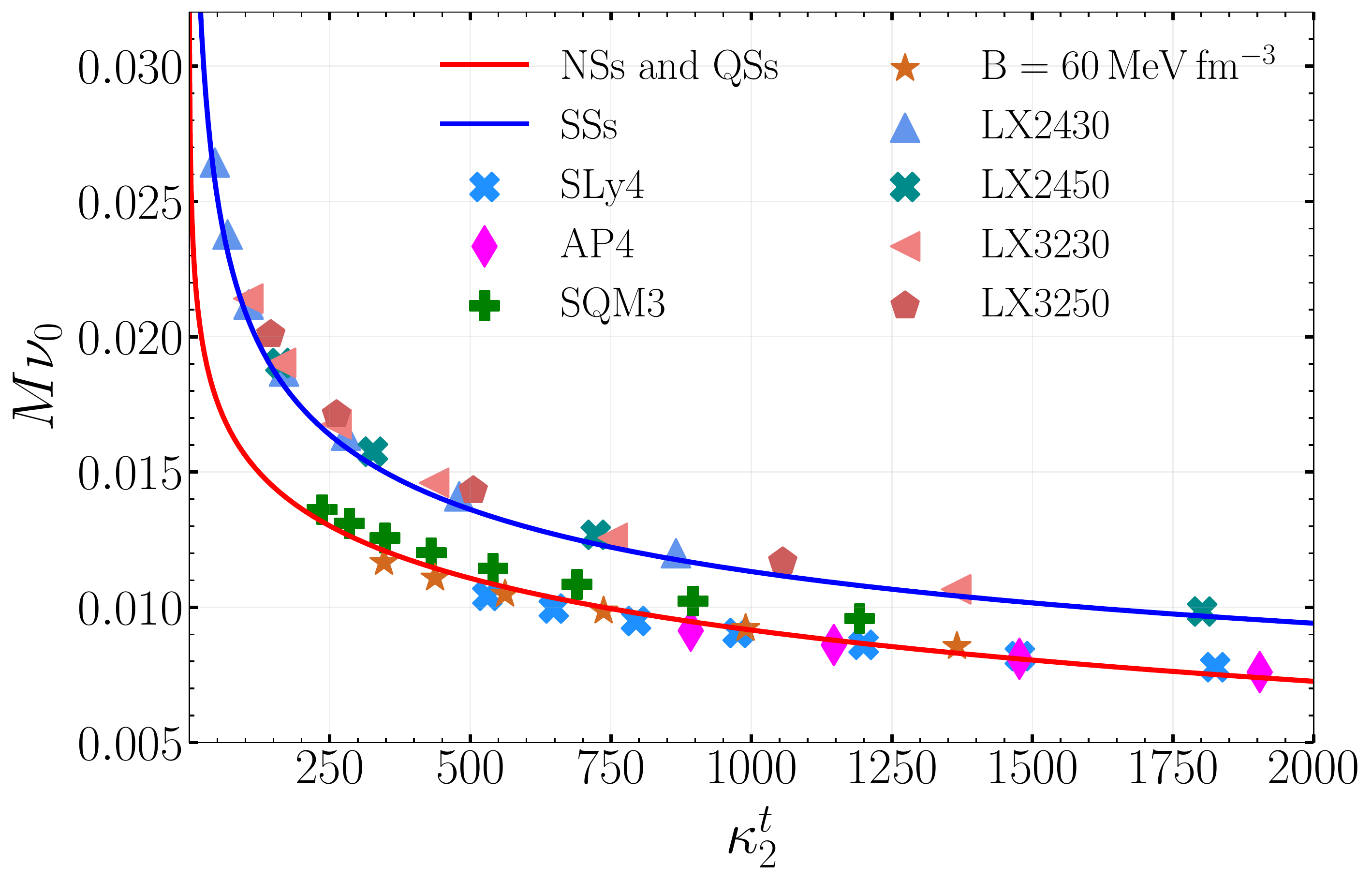}
    \caption{Scaled frequency of the $f$-mode $M \nu_{0}$ as a function of the
    tidal quadrupolar ($l=2$) coupling constant $ \kappa_{2}^{t}$ for NSs, QSs
    and SSs.  The solid line represents the best power law fit in
    $\kappa_{2}^{t}$ to the scaled frequencies of the NSs, QSs and SSs.}
    \label{fig:relation_3}
\end{figure}
%%%%%%%%%%%%%%%%%%%%%%%%%%%%%%%%%%%%%%%%%%%%%%%%%%

Inspired by the universal relation between the dimensionless tidal coupling
constant and the $f$-mode frequency~\citep{Chakravarti:2019sdc}, the relation
between $M\nu_{0}$ and $\kappa_{2}^{t}$ for NSs, QSs, and SSs are shown in
Fig.~\ref{fig:relation_3}.  For NSs and QSs, we find the scaled frequency of the
$f$-mode approximately satisfies the following relation,
%--
\begin{equation}
  M\nu_{0}= 0.184(\kappa_{2}^{t})^{-0.016} - 0.154 \,.
\end{equation}
%--
For SSs, the universal relation is 
%--
\begin{equation}
  M\nu_{0}= 0.071(\kappa_{2}^{t})^{-0.266} \,.
\end{equation}
%--
The universal relations for QSs and SSs will complement that of NSs, and play a
role in GW data analysis~\citep{Dietrich:2017aum}.

%------------------------------------------------------------------------------------
\section{Conclusions}\label{sec: conclusion}
%------------------------------------------------------------------------------------

In this paper,  we use the Lennard-Jones model to describe the EOS of SSs with
two parameters, the number density at the surface of the star $n_{s}$ and the
potential depth $\epsilon$.  Compared to the MIT bag model of QSs, the EOS of
SSs is much stiff due to the nonrelativistic nature of the particles and the
compressed repulsive hardcore at a small intercluster distance.  Following
earlier work~\citep{Lai:2009cn, Gao:2021uus}, we calculate the mass and radius
relation for SSs for different values of $n_{s}$ and $\epsilon$, and find that
the maximal mass of SSs is higher than that of NSs and QSs. This serves as
background solutions for perturbation studies of various oscillation modes.
  
To study radial oscillations of SSs, for the first time we calculate the
frequency of the radial modes for SSs with different combinations of  $n_{s}$
and $\epsilon$.  The results are compared with that of NSs and QSs.  We discover
that radial oscillations of SSs are similar to those of QSs but behave very
differently from those of NSs, especially for stars with low central energy
densities or small masses.  For QSs and SSs, the frequencies of radial
oscillations tend to infinity when the central energy density approaches the
minimal value $\rho_{\rm min}$, which corresponds to the pressure being zero.
This can be understood by approximating the stars in the nonrelativistic regime
and noticing that the adiabatic index $\Gamma$ for SSs and QSs goes to infinity
as the density decreases to its minimal value.

For nonradial oscillations of SSs, we calculate the frequency of the $f$-mode
for $l=2$ component using the Cowling approximation, and obtain the universal
relations between the $f$-mode frequency and other global parameters of the
spherical SSs.  As recently proposed in~\citet{Gao:2021uus}, where the I-Love-Q
universal relations for SSs were studied, the universal relation of the $f$-mode
frequency for SSs is also ready to be used for various purposes in GW
astrophysics involving compact stars. With application to data in the future,
possible constraints can be set on the parameter space of the Lennard-Jones
model, namely the $n_s$-$\epsilon$ plane, using GW observations of the QNMs from
compact stars.

There can be several interesting extensions of our work. First, our study of
nonradial oscillations uses the Cowling approximation~\citep{Cowling:1941},
which considers only the fluid perturbation.  In principle, one should also
allow the spacetime metric to be perturbed, and thus one can obtain QNMs instead
of normal modes.  Next, we would like to further investigate how dynamical tides
affect the frequency of the $f$-mode in compact binary systems.  \Reply{NSs have
certain spins and the rotation rate may reach extreme values, especially for
nascent or remnant objects following a binary merger. From the perspective of
detecting oscillation modes with GWs, the most relevant scenarios are likely to
involve rapidly rotating NSs. An important step in this direction has been
carried out  using perturbation theory in general relativity with the Cowling
approximation \citep{Kruger:2009nw, Gaertig:2010kc, Doneva:2013zqa}. In the next
step, we can study the oscillation modes of rapidly rotating SSs in the Cowling
approximation based on existing work. Recently, \citet{Kruger:2019zuz,
Kruger:2020ykw} managed to calculate the oscillations and instabilities of
relativistic stars using perturbation theory without the Cowling approximation.
The oscillation spectrum, universal relations involving $f$-mode, and the
critical values for the onset of the secular Chandrasekhar-Friedman-Schutz
instability are studied in great detail. Further, \citet{Manoharan:2021wds}
investigated universal relations for binary NS mergers with long-lived remnants.
By considering the oscillations of the rapidly rotating merger remnant, they
proposed an approach to relate the pre-merger tidal deformability to the
effective compactness of the post-merger remnant. Those studies are important to
probe the EOS of NSs with GW asteroseismology. Therefore, to study the
oscillations of rapidly rotating SSs without the Cowling approximation is an
important goal worth pursuing.}

%------------------------------------------------------------------------------------
\section*{Acknowledgements}
%------------------------------------------------------------------------------------

\Reply{We thank Christian Kr\"uger for useful comments.} This work was supported
by the National SKA Program of China (2020SKA0120300, 2020SKA0120100), the
National Natural Science Foundation of China (11975027, 11991053, 11721303), the
National Key R\&D Program of China (2017YFA0402602), the Max Planck Partner
Group Program funded by the Max Planck Society, and the High-Performance
Computing Platform of Peking University. Rui Xu is supported by the Boya
Postdoctoral Fellowship at Peking University.

%%%%%%%%%%%%%%%%%%%%%%%%%%%%%%%%%%%%%%%%%%%%%%%%%%
\section*{Data Availability}
 
The data underlying this paper will be shared on reasonable request to the 
corresponding authors.



\bibliographystyle{mnras}
\bibliography{refs} % if your bibtex file is called example.bib

%\begin{thebibliography}{99}
%\end{thebibliography}

%%%%%%%%%%%%%%%%%%%%%%%%%%%%%%%%%%%%%%%%%%%%%%%%%%

%%%%%%%%%%%%%%%%% APPENDICES %%%%%%%%%%%%%%%%%%%%%

% \appendix

% \section{Some extra material}

% If you want to present additional material which would interrupt the flow of the main paper,
% it can be placed in an Appendix which appears after the list of references.

%%%%%%%%%%%%%%%%%%%%%%%%%%%%%%%%%%%%%%%%%%%%%%%%%%

% Don't change these lines
\bsp	% typesetting comment
\label{lastpage}
\end{document}